\documentclass[10pt, aps, prb, superscriptaddress, noshowkeys, twocolumn, floatfix, preprintnumbers,showpacs]{revtex4-1}

\usepackage{amsmath, amssymb, amsfonts}
\usepackage{graphicx, color}
\usepackage{times}

\usepackage[breaklinks]{hyperref}
\usepackage{bookmark} % get rid of a warning from hyperref
\hypersetup{colorlinks, citecolor = blue, filecolor = blue, linkcolor = blue, urlcolor = blue}  
\newcommand{\ud}{\text{d}}
\renewcommand{\Im}{\,\text{Im}}
\renewcommand{\Re}{\,\text{Re}}
\newcommand{\bk}{{\boldsymbol{k}}}
\newcommand{\bp}{{\boldsymbol{p}}}
\newcommand{\bq}{{\boldsymbol{q}}}
\newcommand{\bK}{{\boldsymbol{K}}}
\newcommand{\bx}{{\boldsymbol{x}}}
\renewcommand{\bm}[1]{\boldsymbol{#1}}
\newcommand{\e}{\hat e}
\newcommand{\h}{\hat h}
\renewcommand{\a}{\hat a}
\renewcommand{\b}{\hat b}

\allowdisplaybreaks

\begin{document}
\title{Optical Response and Ground State of Graphene}

\author{T. Stroucken}
\affiliation{Department of Physics and Material Sciences Center, Philipps University Marburg, Renthof 5, D-35032 Marburg, Germany}

\author{J. H. Gr{\"o}nqvist}
\affiliation{Department of Physics, \AA bo Akademi University, 20500 Turku, Finland}
\affiliation{Department of Physics and Material Sciences Center, Philipps University Marburg, Renthof 5, D-35032 Marburg, Germany}

\author{S.W. Koch}
\affiliation{Department of Physics and Material Sciences Center, Philipps University Marburg, Renthof 5, D-35032 Marburg, Germany}

\begin{abstract}
The optical response and the ground state of graphene and graphene-like systems are determined selfconsistently. 
Deriving equations of motion for the basic variables, graphene Bloch equations are introduced and combined 
with a variational Ansatz for the ground state.
Within the Hartree--Fock approximation, this approach reproduces the gap equation for the ground state. 
The results show that the Coulomb interaction drastically influences the optical response of graphene and 
introduces an extremely sensitive dependency on the dielectric environment via screening.  
Regarding the effective fine-structure constant as control parameter, a transition from a semimetal to 
an excitonic insulator is predicted as soon as the effective graphene fine-structure constant exceeds a 
value of roughly 0.5. 
Above this critical value, the computed optical spectra exhibit a pseudogap and several bright $p$-like excitonic resonances.

\end{abstract}
\date{\today}


\maketitle

\section{Introduction}

The research interest in graphene has increased dramatically since its first isolation by Geim and co-workers in 2004. 
In particular its highly unusual electronic and optical properties has led to a cascade of both 
theoretical and experimental investigations, exploiting its fundamental underlying physics as well as 
potential applications in electronic and optoelectronic devices. 

The key for understanding the unique electronic properties of graphene is its exotic band structure that differs 
substantially from most other condensed matter systems. Based on a tight-binding (TB) model, Wallace predicted 
1947\cite{Wallace1947} an electronic single-particle spectrum exhibiting two distinct crossing points. 
In the vicinity of these so called Dirac points, the dispersion is a cone similar to the light cone in relativistic 
mechanics, with the Fermi velocity $v_F$ replacing the speed of light.
The occurrence of the cones results from the symmetry between the two equivalent sublattices that build the honeycomb 
lattice. The sublattice wave functions can be combined into a pseudo-spinor which then obeys the ultra-relativistic Dirac equation. 
Hence, from a QED point of view, electronic excitations close to the Dirac points of graphene can be considered as charged, 
massless, chiral fermions. From a condensed matter point of view, as the density of states vanishes at the Dirac points, graphene 
can be considered either as a semimetal or a vanishing gap semiconductor.

However, as the tight binding Hamiltonian neglects many body interactions completely, 
the role of the electron--electron Coulomb interaction is still not well understood and the subject of ongoing 
research.\cite{Gonzalez99, Sheehy2007, Fritz2008, Sinner2010, Khveshchenko2001a, Khveshchenko2006, Juricic2009, Drut2009a,
Drut2009b, Yang2009, Malic2010, Guclu2010, Reed2010, Sabio2010a, Sabio2010b, Gamayun2009, Gamayun2010, Wang2010, Wang2011,
Herbut2006, Herbut2008, Shytov2007b, Pereira2007, CastroNeto2009b, Heinz2011, Chae2011}
As a convenient measure of the relative importance of the Coulomb interaction, one can use the 
effective fine-structure constant $\alpha_\text{G} = e^2/ \epsilon \hbar v_F$,
where $\epsilon$ is the effective background dielectric constant. For freestanding graphene in vacuum 
$\alpha_\text{G} \approx 2.41$, indicating prominent Coulomb interaction effects. 

Generally, one can distinguish between Coulomb modifications of the electronic ground state and Coulombic signatures 
in the excitation properties. From strongly correlated systems it is known that the Coulomb interaction can induce 
a transition from a semimetal to a Mott insulator where, unlike in conventional semiconductors, the gap results from 
the electron--electron rather than the electron--ion interactions. In particular, the electron exchange interaction 
has been identified as the dominant mechanism responsible for the opening of a gap.\cite{Engel2009}

Methodically, most theoretical treatments of the ground state of a many body systems either use highly 
simplified model Hamiltonians or they rely on either perturbative or variational approaches.
For graphene, perturbative studies based on a renormalization-group analysis predict a logarithmic divergence of the Fermi 
velocity, stabilizing the semimetallic ground state.\cite{Gonzalez99, Sheehy2007, Fritz2008, Sinner2010}
However, nonperturbative methods yield a semimetal to insulator transition at sufficiently high coupling
strengths\cite{Khveshchenko2001a, Khveshchenko2006, Juricic2009, Drut2009a, Drut2009b} where predicted critical 
values for $\alpha_\text{G}$ range from 0.5 to 1.5. 
In particular, the possibility of an excitonic condensate has been explored.\cite{Sabio2010b, Gamayun2010}

From semiconductor physics, it is known that the excitation spectrum in the vicinity of the fundamental band
gap is dominated by Coulomb bound electron--hole pairs leading to the appearance of excitonic resonances in the absorption spectra. 
Mathematically, the excitons obey a hydrogen-like Schr\"odinger equation, which is known as Wannier equation.\cite{HaugKoch}
The exciton binding energy in typical semiconductors is in the range of 1 to 100 meV, i.e. much smaller than the 
hydrogen ground binding energy. 
However, bound states exist even in the presence of strong static (background) screening which preserves the 
long-range $1/r$ Coulomb tail. In metals, the highly mobile carriers effectively screen this long ranged part of the 
Coulomb interaction with the consequence that excitonic effects are of minor importance. 
In graphene, the existence of bound pair states and the importance of screening effects are subjects of 
current research. Since real massless particles in nature are neutral, the Dirac two-body problem has become a topic of 
interest only within the graphene research and, until recently, it has not even been clear if bound Dirac pairs 
exist.\cite{Wang2010, Gronqvist2011}

Experimentally, an important tool to study microscopic processes in  many-body systems is optical spectroscopy, providing 
information both on the system ground state and the excitation properties. 
Theoretically, several methods to model the optical response of a quantum mechanical many-body system are well established,
e.g. the density matrix approach, non-equilibrium Green functions, or the systematic cluster expansion approach.\cite{Kira2006}
In general, these methods rely on the knowledge of the initial state, which is usually the ground state. 

In this paper, we extend the microscopic approaches and develop a framework that allows us to determine the ground state 
and the optical response of graphene and graphene-like systems on the same level of approximation. Our method combines the 
equations of motion with a variational approach. Within the Hartree--Fock approximation, we obtain gap equations that 
determine a Bogoliubov ground state.\cite{Sabio2010b, Gamayun2010}

With this initial state, we then calculate the linear optical response. The resulting spectra show several bright excitonic 
resonances below the pseudo-gap.

The work is organized as follows. In Section \ref{sec:Hamiltonian}, we present the Hamiltonian within the tight-binding 
approximation. Using this basis, we derive the equations of motion for the dynamical quantities of interest and discuss 
the role of the Coulomb interaction. 
Similar equations have been derived previously for carbon nanotubes  \cite{Hirtschulz2008, Gronqvist2010}.
In section \ref{sec:Groundstate}, we apply the variational principle to the total 
Hamiltonian and impose the equations of motion as constraint to obtain a stationary ground state. The resulting gap 
equations are presented and analyzed numerically. In Section \ref{sec:GBE}, we derive the graphene Bloch
equations and compute the linear optical response.

\section{The Hamiltonian and Equations of Motion} \label{sec:Hamiltonian}

\begin{figure}
\centering{\includegraphics[width=8.5cm]{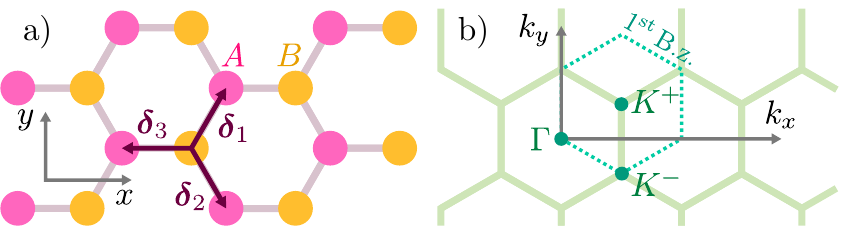}}
\caption{The lattice and the reciprocal lattice.}
\label{fig:lattice}
\end{figure}

The system Hamiltonian for a single graphene sheet interacting with a classical light field can be written as 
\[ \hat{H} = \hat{H}_0 + \hat{H}_I + \hat{H}_C .\]
Here,
\begin{eqnarray}
\hat{H}_0 
&=& \int\ud^3x \ \hat\psi^\dagger(\bx) \biggl\{\frac{{\bm p}^2}{2m_0} + \sum_{\bm R_A, \bm R_B} V(\bx - \bm R)\biggr\}
 \ \hat\psi(\bx),\quad
\end{eqnarray}
describes the motion of the electrons of mass $m_0$ in the periodic lattice potential,
$V(\bx-\bm R)$ is the effective core potential of the 
carbon atom located at $\bm R$, and
 $\{\bm R_A \}$ and $\{\bm R_B\}$
are the coordinates of the carbon atoms on each sublattice (See Fig. \ref{fig:lattice}). 
\begin{eqnarray}
\hat{H}_C 
&=& \frac{1}{2} \int \ud^3x \int\ud^3x' \ 
\hat{\psi}^\dagger(\bx) \hat{\psi}^\dagger(\bx') V(\bx - \bx') \hat{\psi}(\bx') \hat{\psi}(\bx)\nonumber\\
\end{eqnarray}
describes the electron--electron interaction via the Coulomb potential $V(\bx-\bx') = e^2/\epsilon|\bx-\bx'|$
where
$\epsilon$ is the dielectric constant of the environmental medium.
\begin{eqnarray}
\label{HI}
\hat{H}_I 
&=& -\frac{e}{2m_0 c} \int\ud^3x  \nonumber\\
&& \hat{\psi}^\dagger(\bx) \biggl\{ \bm p \cdot \bm A (\bx) + \bm A(\bx) \cdot {\bm p} 
- \frac{e}{c}{\bm A}^2 \biggr\} \hat{\psi}(x) \quad
\end{eqnarray}
describes the light--matter interaction within the minimal coupling substitution and ${\bm A}$  is the vector potential 
for the optical field.

The interaction Hamiltonian couples the dynamics for the vector potential to the expectation value of the particle current
that can be calculated from the Heisenberg equation of motion, 
\[
i\hbar\frac{\ud}{\ud t}\langle \hat O\rangle =
\langle \left[ \hat O, \hat H \right] \rangle + i\hbar\langle\frac{\partial}{\partial t} \hat O\rangle ,
\]
for the relevant operator $\hat O$, yielding a set of coupled differential equations.
As is well known, the two-particle Coulomb interaction couples the dynamics for the $N$-particle expectation values to those of 
the $N+1$-particle expectation values, which is known as the hierarchy problem. In order to achieve a closed set of equations, 
this hierarchy must be truncated. A systematic truncation scheme is provided by the cluster expansion, which has been proven 
to work quit well under many different excitation conditions.\cite{Kira2006}

In principle, the system dynamics can be described by solving the resulting coupled set of differential equations starting 
from a predetermined initial state. In a typical experimental setup, the system is excited from the ground state and the 
response to an externally applied field is measured. 
Thus, to analyze such experiments theoretically, it is crucial to have an adequate description of the ground state.
In semiconductor physics, band structure calculations usually provide a suitable basis to expand the field operators 
and a good approximation for the ground state. However, this may not be the case in systems where strong carrier--carrier 
Coulomb interactions influence the ground state properties.

In the following, we therefore follow an approach where we treat $H_0$ within the tight-binding approximation and use the 
resulting eigenfunctions as basis to represent the total Hamiltonian. Within this basis, we then derive the equations of 
motion on the singlet level, which is equivalent to the time-dependent Hartree--Fock approximation. 
To determine the ground state on the same level of approximation, we apply the variational approach for the system energy and 
impose the equations of motion as constraints to guarantee stationarity of the ground state. 

\subsection{Tight-binding Hamiltonian}

Following the tight-binding approach, we expand the field operators in terms of the carbon wave functions,
\begin{eqnarray}
\hat{\psi}(\bx)
&=& \frac{1}{\sqrt{N}} \sum_{\bk ,\bm R_A} e^{i\bk \cdot \bm R_A} \phi(\bx-\bm R_A) \a_{\bk} \nonumber\\
&+& \frac{1}{\sqrt{N}} \sum_{\bk, \bm R_B} e^{i\bk \cdot \bm R_B} \phi(\bx-\bm R_B) \b_{\bk} \nonumber\\
&=& \hat{\psi}_A(\bx) + \hat{\psi}_B(\bx).\label{fieldoperator}
\end{eqnarray}
Here, $\hat{a}_\bk$ ($\hat{b}_\bk$) annihilates a particle in the state $\left\{\bk\right\}$  on 
the sublattice $A$ ($B$), and $
\phi(\bm r)=  r\, e^{-r/2d} \cos\vartheta / \sqrt{{32\pi d^5}}$ is the carbon $2 p_z$ orbital 
responsible for the optical and electronic electronic properties of graphene.
The parameter $d = a_B/Z_{{\rm eff}}$ controls the effective spreading of the carbon wave functions.
  
\begin{figure}
\centering{\includegraphics[width=8.5cm]{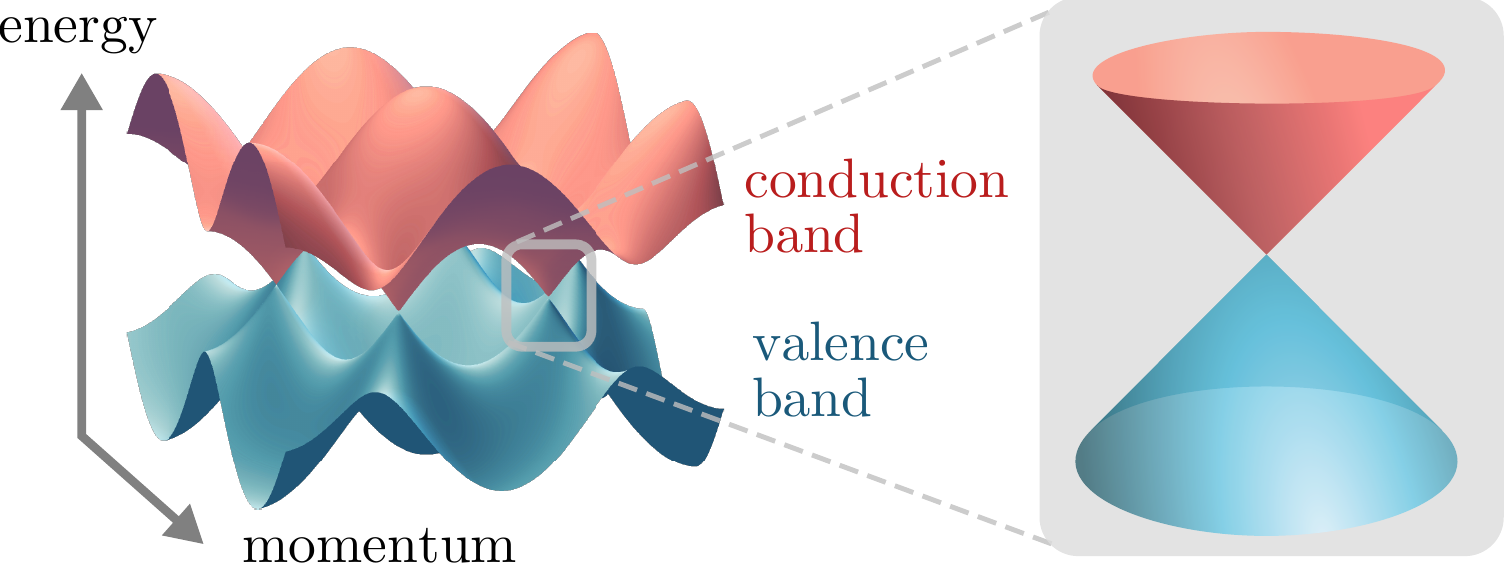}}
\caption{Schematic plot of the $\pi$-valence and the $\pi^*$-conduction band structure of graphene 
according to the tight-binding model}
\label{fig:bandstructure}
\end{figure}

Inserting Eq. (\ref{fieldoperator}) into the Hamiltonian and taking into account nearest neighbor hopping only, 
one obtains for the single-particle part
\begin{equation} \label{eq:TB0}
\hat{H}_0 = \sum_{\bk} E_F \left( \a^\dagger_\bk \a_\bk + \b^\dagger_\bk \b_\bk \right)
+ \gamma f(\bk) \a^\dagger_\bk \b_\bk + \gamma f^*(\bk) \b^\dagger_\bk \a_\bk .
\end{equation}
Here, $E_F$is the expectation value of $\hat H_0$, $\gamma$ is the matrix element describing 
hopping processes between neighboring carbon atoms,
${\bm \delta}_1 = \frac{a}{2} (1,\sqrt{3})$,
${\bm \delta}_2 = \frac{a}{2} (1,-\sqrt{3})$, and
${\bm \delta}_3 = (-a,0)$
are the vectors connecting any carbon atom to its three next neighbors (see Fig. \ref{fig:lattice}), 
$a$ is the nearest neighbor distance and
\[f({\bk}) = \sum_{i=1}^3 e^{i{\bk}\cdot {\bm \delta}_i} \] 
is a function that depends on the lattice symmetry properties only.
Diagonalization of $\hat{H}_0$ yields the TB single-particle band structure
\begin{equation}
E^{c/\nu}_{\bk} = E_F \pm \gamma|f({\bk})| \label{tbdispersion}
\end{equation}
shown if Fig. \ref{fig:bandstructure}.

Within the tight binding approximation, the shape of the bands depends on the lattice symmetry only.
The valence and conduction bands are symmetric and touch each other at the two nonequivalent points 
${\bm K}^\pm = (2\pi/3a, \pm 2\pi/3\sqrt{3}a)$ of the Brillouin zone, where the function $f({\bk})=0$. 
Since the ground state of Eq. (\ref{eq:TB0}) has a completely filled valence and an empty conduction band, 
the nodes occur exactly at the Fermi level $E_F$.
In the vicinity of these nodes, a first order Taylor expansion gives
\[
f(\bm K^\pm + \bk) = -\frac{3a}{2} e^{-i\pi/6} (k_x \pm ik_y).
\]
Thus, the dispersion is a cone with linear coefficient $3\gamma a/2\equiv\hbar v_F$,
similar to the light cone in relativistic mechanics. Consequently the electrons behave like 
massless Dirac fermions with the Fermi velocity replacing the speed of light. 
Both experimental and theoretical data give a value of approximately $10^6$ m/s for the Fermi velocity.

In the following, we will treat $\hat{H}_0$ within the TB approximation.
Within the band structure picture, the Hamiltonian is represented in terms of the 
electron and hole operators
\begin{eqnarray}
\e_{\bk }          &=& \frac{1}{\sqrt{2}} \left(  \a_{\bk } + \tilde f({\bk}) \b_{\bk }\right)\\
\h^\dagger_{-\bk } &=& \frac{1}{\sqrt{2}} \left(- \a_{\bk } + \tilde f({\bk}) \b_{\bk }\right),
\end{eqnarray}
with $\tilde f({\bk}) = f({\bk})/|f({\bk})|$, yielding
\begin{eqnarray}
\hat{H}_0 
&=& \sum_{\bk} \gamma|f(\bk)| \left( \e^\dagger_\bk \e_\bk + \h^\dagger_{-\bk} \h_{-\bk} \right)\nonumber\\
&+& \sum_{\bk} E_F \left( \e^\dagger_\bk  \e_\bk - \h^\dagger_\bk \h_\bk \right)
+\sum_{\bk }\left( E_F-\gamma|f({\bk})|\right). \quad
\end{eqnarray}
Here, the second term only contributes if the electron and hole symmetry is broken, e.g. by doping, and the last term 
represents the energy of the filled valence band. Since this term is constant, it can be omitted.
Taking the tight-binding ground state as a reference, the energy with respect to this reference 
is obtained by normal ordering the Hamiltonian within the electron--hole picture.

Important Coulomb contributions arise from scattering processes where each electron 
remains on its specific sublattice, defining the generic matrix element
\[
V(\bq)=\frac{2\pi e^2}{\epsilon q}F(qd)
\]
that depends on the momentum transfer only and is given by the 2D bare Coulomb potential modified by the 
background dielectric constant $\epsilon$ and a form factor $F(qd)$. The form factor
\[
F(qd) = \int \ud^3r \int \ud^3r'  e^{i \bq \cdot( \bm\rho - \bm\rho')}  e^{-q|z-z'|} |\phi(\bm r)|^2|\phi(\bm r')|^2 
\]
with \hbox{$F(0)=1$} 
results from the finite extension of the carbonic $p_z$-orbitals perpendicular to the plane and decreases monotonically with $q$.
Hence, the finite value of $d=a_B/Z_{\rm eff}$  can be interpreted as the effective thickness of the graphene sheet. 
As an intrinsic length scale, it fixes the graphene energy unit $E_0=\hbar v_F/d$ and is crucial for obtaining 
finite values for the exciton binding energy.\cite{Gronqvist2011} 
The Coulomb matrix elements for the processes where at least one electron is scattered from one sublattice to
the other are much smaller and vanish exactly at the Dirac points. These will be neglected.

\begin{figure}
\centering{\includegraphics[width=8.5cm]{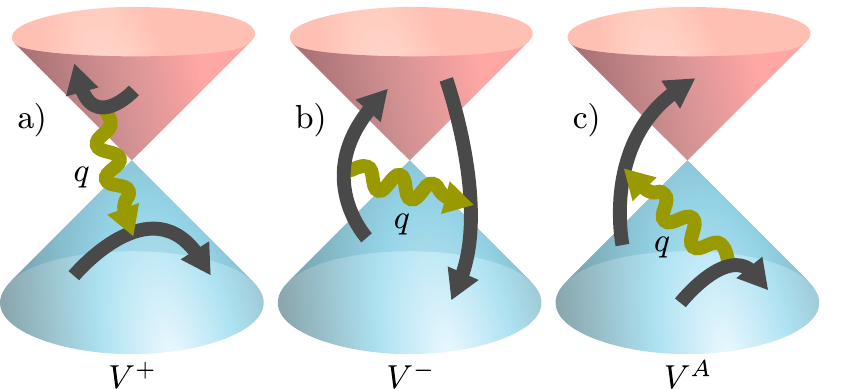}}
\caption{Schematic representation of the Coulomb matrix elements}
\label{fig:Coul}
\end{figure}

Expanding the field operators in terms of the electron and hole wave functions produces $2^4$ different contributions, some of 
which describe equivalent processes. The $10$ nonequivalent normally ordered contributions to the Coulomb interaction
\begin{eqnarray}
H_C
&=& \frac{1}{2} \sum_{\bq\bk\bk'} V^+_{\bk\bk'}(\bq)
\e^\dagger_{\bk+\bq} \e^\dagger_{\bk'-\bq} \e_{\bk'} \e_{\bk} \nonumber\\
&+& \frac{1}{2} \sum_{\bq\bk\bk'} V^+_{\bk\bk'}(\bq)
\h^\dagger_{-\bk-\bq} \h^\dagger_{-\bk'+\bq} \h_{-\bk'} \h_{-\bk} \nonumber\\
&-& \sum_{\bq \bk\bk'} V^+_{\bk\bk'}(\bq)
\e^\dagger_{\bk+\bq} \h^\dagger_{-\bk'-\bq} \h_{-\bk'} \e_{\bk} \nonumber\\
&+& \sum_{\bq\bk\bk'} V^-_{\bk\bk'}(\bq)
\e^\dagger_{\bk+\bq} \h^\dagger_{-\bk} \h_{-\bk'+\bq} \e_{\bk'} \nonumber\\
&+& \frac{1}{2} \sum_{\bq\bk\bk'} V^-_{\bk\bk'}(\bq)
\e^\dagger_{\bk+\bq}\e^\dagger_{\bk'-\bq} \h^\dagger_{-\bk'} \h^\dagger_{-\bk} \nonumber\\
&+& \frac{1}{2} \sum_{\bq\bk\bk'} V^-_{\bk\bk'}(\bq)
\h_{-\bk-\bq} \h_{-\bk'+\bq} \e_{\bk'} \e_{\bk} \nonumber\\
&-& \sum_{\bq\bk\bk'} V^A_{\bk\bk'}(\bq)
\e^\dagger_{\bk+\bq} \h^\dagger_{-\bk'} \e^\dagger_{\bk'-\bq} \e_{\bk} \nonumber\\
&-& \sum_{\bq\bk\bk'} V^A_{\bk\bk'}(\bq)
\e^\dagger_{\bk+\bq} \e_{\bk'} \h_{-\bk'+\bq} \e_{\bk} \nonumber\\
&+& \sum_{\bq\bk\bk'} V^A_{\bk\bk'}(\bq)
\h^\dagger_{-\bk} \h_{-\bk'+\bq} \h_{-\bk-\bq} \e_{\bk'} \nonumber\\
&+& \sum_{\bq\bk\bk'} V^A_{\bk\bk'}(\bq)
\e^\dagger_{\bk'-\bq} \h^\dagger_{-\bk} \h^\dagger_{-\bk'} \h_{-\bk-\bq},\label{HCoulomb}
\end{eqnarray}
 have three different matrix elements
\begin{eqnarray}
V_{\bk\bk'}^\pm(\bq)&=& \frac{1}{4}V(q)\left(1\pm \tilde f^*({\bk+\bq})\tilde f({\bk})\right)\nonumber\\
&\times&\left(1\pm \tilde f^*({\bk'-\bq})\tilde f({\bk'})\right)\\
V^A_{\bk\bk'}(\bq)&=& \frac{1}{4}V(q)
\left(\tilde f^*({\bk+\bq})\tilde f({\bk})-1\right)\nonumber\\
&\times&
\left(1+\tilde f^*({\bk'-\bq})\tilde f({\bk'})\right)
\end{eqnarray}
that not only depend on the momentum transfer but also on the momenta of the involved scattering particles. Only within the 
linear approximation, translational invariance is recovered and the matrix elements 
with $\bq=\bk'-\bk$, relevant for the Hartree--Fock approximation, are given by
\begin{eqnarray}
 V^\pm(\bk-\bk')&=& 
\frac{1}{2} V(|\bk - \bk'|) \left( 1 \pm \cos(\theta-\theta') \right)\\
V^A(\bk-\bk')& = &\pm\frac{i}{2} V(|\bk - \bk'|) \sin(\theta-\theta').
\end{eqnarray}
The scattering processes characterized by the different matrix elements are schematically presented in Fig. \ref{fig:Coul}.
The repulsive e--e and h--h and the attractive e--h scattering where both particles remain in their initial bands 
are described by  $V^+$ (Fig. \ref{fig:Coul}a).
The corresponding processes, where both particles change their bands, are described by 
$V^-$ (Fig. \ref{fig:Coul}b). Both $V^+$ and $V^-$ are symmetric with respect to $\bk-\bk'$ and equal for both 
Dirac points.
Auger processes, where an electron--hole pair is created or annihilated under simultaneous energy and momentum transfer 
to other quasiparticles, are not ruled out by energy and momentum conservation and described by the matrix elements 
$V^A$ (Fig. \ref{fig:Coul}c). These matrix elements are antisymmetric with respect to $\bk-\bk'$ and have opposite sign 
for the two distinct Dirac points. As usual, processes with zero momentum transfer describe a divergent self-interaction 
that cancels with the electron--ion and ion--ion interaction in the jellium limit and are explicitly subtracted from the 
Coulomb Hamiltonian.

When computing the optical response, the light--matter interaction is often treated within the dipole approximation. 
In crystalline solids,
the dipole approximation can be obtained by a multipole expansion and subsequent coarse graining over an 
elementary lattice cell, taking only monopole and dipole contributions into account. In typical direct-gap semiconductors,
the distinct symmetry properties of the 
valence and conduction bands allow us to associate monopole contributions to intraband- and dipole 
contributions to interband transitions, respectively. 
As inter- and intraband transitions involve very different energy scales in a large-gap semiconductor,
an optical field couples only to interband transitions while intraband transitions are in the teraherz range.

In graphene, the situation is quite different. As both the $\pi$-valence and $\pi^*$-conduction band are constructed 
from $p_z$ atomic orbitals, they have the same angular momentum quantum numbers and monopole and dipole 
contributions do not distinguish between different bands.
Hence, dipole transitions necessarily involve a superposition of ${\bk}$-states 
required to build a $p$-like collective state.  Additionally,
due to the vanishing gap at the Dirac points, inter- and intraband transitions take place on the same energy scale.

To derive the correct interaction Hamiltonian, we start from Eq. (\ref{HI}),
apply the Coulomb gauge ${\bm \nabla}\cdot{\bm A}=0$,  and
expand the field operators in terms of the tight binding wave functions.
Making the assumption that the vector potential varies slowly on the length scale of the lattice constant $a$ and 
sheet thickness $d$, one finds
\begin{eqnarray}
\hat{H}_I &=& - \frac{e}{m_0c} \sum_{\bk} {\bm A}\cdot\left( \hbar\bk -\frac{e}{2c}{\bm A}\right)
\left(\a^\dagger_{\bk} \a_{\bk} + \b^\dagger_{\bk} \b_{\bk} \right) \nonumber\\
&-& \frac{e}{m_0c} \sum_{\bk} \left( \bm\pi({\bk})\cdot{\bm A} \  \a^\dagger_{\bk} \b_{\bk} 
+ \text{h.c.} \right) \label{HILattice}
\end{eqnarray}
where $\bm{A}$ denotes the field at the position of the graphene sheet, $z=0$.
The  first term in the first line of Eq. (\ref{HILattice}) describes the interaction of the vector potential with the directed current
$\hbar{\bm k} \hat\rho_{\bk}/m_0$, where $\hat\rho_{\bk}=-e(\a^\dagger_{\bk} \a_{\bk} + \b^\dagger_{\bk} \b_{\bk} ) $ 
is the charge density. 
This term vanishes because of parity.
The second term in the first line of Eq. (\ref{HILattice}) is proportional to the total  electron density $n_0$
of the $\pi$-band, yielding a Drude-like response. 
The remaining term describes true quantum mechanical transitions, whose strength is determined by the
optical matrix element
\begin{equation}
{\bm \pi}(\bk) = \sum_{i}{\rm e}^{i{\bk}\cdot{\bm \delta}_i}\int d^3 x \phi^*({\bm r}){\bm p}
\phi({\bm r}-{\bm\delta_i})=-i\frac{2M}{3a^2} \nabla_{\bk} f(\bk)
\end{equation}
and
\begin{equation}
M=\int \ud^3 x \phi^*(\bm r) \bm\delta_i \cdot \bm p \ \phi(\bm r - \bm\delta_i).
\end{equation}
On-site momentum matrix elements vanish because of parity and hence,
in graphene, optical transitions involve inter sublattice hopping processes.
Within the linear approximation, the dipole matrix elements in the vicinity of the Dirac points are given by
\[
\bm \pi_\pm(\bk) =  i\frac{\sqrt{2}M}{a} e^{-i\pi/6} \bm u_\pm = -|\bm \pi| e^{-i\pi/6} \bm u_\pm
\]
with $\bm u_\pm=(\bm e_x \pm i\bm e_y)/\sqrt{2}$, showing that the two degenerate $\bm K$ 
points couple to circular polarization components of the optical field.

Transforming the light--matter interaction Hamiltonian into the 
electron--hole picture yields within the linear approximation,
\begin{eqnarray}
\hat{H}^{\left[p\right]}_I
&=& \frac{e|\bm\pi|}{m_0c} \sum_{\bk} A^\pm \cos\theta
\left( \e^\dagger_{\bk} \e_{\bk} + \h^\dagger_{-\bk} \h_{-\bk}-1 \right) \nonumber\\
&\pm& i\frac{e|\bm\pi|}{m_0c} \sum_{\bk} A^\pm \sin\theta
\left( \e^\dagger_{\bk} \h^\dagger_{-\bk} - \h_{-\bk} \e_{ \bk} \right) \label{eq:HI}
\end{eqnarray}
where the $\pm$ sign refers to the distinct Dirac points and circular polarization components respectively.
Eq. (\ref{eq:HI}) clearly shows that optical excitations not only involve interband, but also intraband transitions.
Furthermore, the angle dependence in the light--matter Hamiltonian assures that only $p$-like states couple to 
an external optical field.

\begin{figure}
\centering{\includegraphics[width=8.5cm]{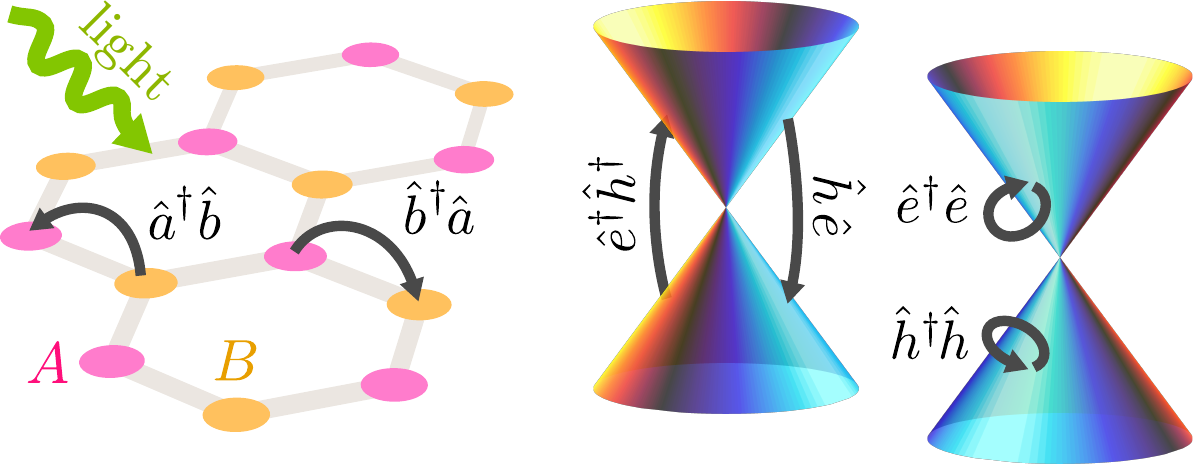}}
\caption{Schematic representation of the light--matter interaction}
\label{fig:optical}
\end{figure}

\subsection{Equations of Motion}

In this section, we use the basic single-particle expectation values
\begin{eqnarray}
 f_{\bk} &=& \langle \e^\dagger_{\bk} \e_{\bk} \rangle = \langle \h^\dagger_{-\bk} \h_{-\bk}\rangle \nonumber\\
 P_{\bk} &=& \langle \h_{-\bk} \e_{\bk}\rangle
\end{eqnarray}
as dynamical variables and derive their equations of motion within the time-dependent Hartree--Fock approximation. 
Due to the band symmetry, the electron and hole populations are equal. Furthermore, as the
two equations of motion for the two distinct $\bK$ points are related by the parity transformation
$\bk\rightarrow -\bk,\ A^\pm\rightarrow A^\mp$, we restrict our discussion to a single Dirac point.

Evaluating the commutators in the Heisenberg equations of motion, we obtain the closed set of differential equations
\begin{equation} \label{GBEpol}
i\hbar \frac{\ud}{\ud t} P_{\bk} 
= 2 \Sigma_\bk P_\bk - (1-2f_{\bk}) \Omega_{\bk}
- \left. i\hbar\frac{\ud}{\ud t} P_{\bk} \right|_{\text{coll}}
\end{equation}
\begin{equation} \label{GBEpop}
\hbar \frac{\ud}{\ud t} f_{\bk} 
= -2\Im \left[ P^*_{\bk} \Omega_{\bk} \right]
-\left. \hbar\frac{\ud}{\ud t} f_{\bk} \right|_{\text{coll}} .
\end{equation}
Here, we introduced the notation
\begin{eqnarray}
 \Sigma_\bk
&=& \hbar v_F k - \sum_{\bk'} \bigl[ V^+(\bk-\bk') - V^-(\bk-\bk') \Bigr] f_{\bk'} \nonumber\\
&-& i\sum_{\bk'} V^A(\bk-\bk') \Im P_{\bk'} 
+\frac{e |\bm \pi|}{m_0 c} A^\pm \cos\theta \nonumber\\
&\equiv&
\Sigma^r_\bk\left[f_\bk\right] +\Delta\Sigma_\bk\left[P_\bk,P^*_\bk,A\right] \label{erenorm}\\
\Omega_{\bk}
&=& \sum_{\bk'} V^+(\bk-\bk') P_{\bk'} + V^-(\bk-\bk') P^*_{\bk'} \nonumber \\
&\mp& i \frac{e|\bm \pi|}{m_0 c} A^\pm \sin\theta - 2 \sum_{\bk'} V^A(\bk-\bk') f_{\bk'} \nonumber\\
&\equiv&\Omega^R_{\bk}\left[P_\bk,P^*_\bk,A\right]-i\Delta\Omega_\bk\left[f_\bk\right] .
\label{rabi}
\end{eqnarray}
In Eqs. (\ref{GBEpol}) and (\ref{GBEpop}), the terms $\ud / \ud t |_\text{coll}$ refer to incoherent scattering contributions 
beyond the Hartree--Fock approximation. To solve the coupled equations, we have to supply the appropriate initial and 
boundary conditions. 

The generalized Rabi energy $\Omega_\bk$ consists of two contributions.
$\Omega^R_\bk$ contains the external field and the internal field of the polarization $P_\bk$ and differs
from the standard Rabi frequency known from semiconductors only by the anisotropy resulting from the chiral nature of the 
Dirac electrons. The additional part $\Delta\Omega$
contains contributions
from the populations $f_\bk$  via the Auger processes. 
Similarly, the generalized renormalized energy $\Sigma_\bk$  
has not only terms due to the populations contained in $\Sigma^r$,  but also Auger contributions from the 
polarizations and an energy renormalization due to the 
external field in $\Delta\Sigma_\bk$. For large gap semiconductors, these contributions can be neglected because of energy 
and momentum conservation. However, for systems with small or vanishing gap, as in graphene, the basic quantities $f$ and 
$P$ are inherently mixed, due to the Auger terms.

Close to the $\bK^\pm$ point, the Auger contributions to $\Omega$ and $\Sigma$ can be written as
\begin{eqnarray}
\Delta\Omega_\bk&=& \pm  \sum_{\bk'} V(\bk-\bk') \sin(\theta-\theta') f_{\bk'}\nonumber\\
\Delta\Sigma_\bk& =&  \pm \sum_{\bk'} V(\bk-\bk') \sin(\theta-\theta') \Im P_{\bk'} ,\nonumber
\end{eqnarray}
respectively. For isotropic densities $f$ and $P$, the angle integration will give zero, and the Auger 
contributions vanish. The light--matter terms, however, are in themselves anisotropic. The light will therefore 
excite angular-dependent densities for which the Auger contributions are finite.

In semiconductor physics, the transverse optical field directly couples only to the nondiagonal expectation values $P$ and $P^*$, 
which has the meaning of an optical polarization and can be probed by optical spectroscopy.
Unlike in graphene, the diagonal populations $f_{\bk}$ affect optical semiconductor spectra only indirectly
via their coupling to the polarization in the semiconductor Bloch equations (SBE).\cite{HaugKoch}
As the linear coupling to the optical field results in a radiative decay of any macroscopic polarization on
a picosecond time scale,\cite{Stroucken1996} a (meta-)stable initial state may be prepared to  contain incoherent 
diagonal populations but no macroscopic polarizations. In particular, choosing the initial $f_{\bk}=0$ models excitation 
from the ground state. Expanding the solution of the SBE into powers of the exciting fields 
gives the well known series that contains only the odd powers for the polarization and the even powers of the 
optical field for the populations.\cite{MeierThomasKoch}
Hence, for weak optical fields, the equations for the polarizations and populations can be solved iteratively. 
The homogeneous part of the polarization equation produces the generalized Wannier equation with the Pauli blocking 
prefactor $(1-2f)$. This homogeneous part is responsible for the formation of bound excitons at low densities, that 
appear as resonances of the excited system, energetically below the band gap. Inserting increasing amounts of incoherent initial
populations leads to a gradual bleaching of excitonic resonances  in the absorption spectra and finally produces gain once 
inversion is reached.

A similar behavior might be expected for the case of graphene. However, the interpretation of the 
single-particle energy renormalization as "band edge" renormalization is problematically in a gapless system. Also,
bound exciton solutions of the Wannier equation would predict resonances at negative energies, which is clearly unphysical.
This pathological behavior would show up for any incoherent initial population that allows for bound-state solutions of 
the generalized Wannier equation. Consequently, such a population cannot correspond to a physically meaningful initial 
state of the system. 

In particular, if the linear Wannier equation
\begin{eqnarray} 
2\hbar v_F k \,\phi_{\lambda\bk}
&-&\sum_{\bk '}
\left[ V^+(\bk-\bk ') \phi_{\lambda \bk'} + V^-(\bk-\bk ') \phi^*_{\lambda \bk'} \right] \nonumber\\
&=& E_{\lambda} \, \phi_{\lambda\bk} \, .\label{Wannierequation}
\end{eqnarray}
has bound state solutions, the tight-binding ground state with $f_{\bk}=P_{\bk}=0$ cannot correspond to the ground state of 
the Coulomb-interacting system. The linear Wannier equation has been analyzed in  Ref. \onlinecite{Gronqvist2011}. 
Here, it has been shown that bound-state solutions indeed exist for an effective graphene fine-structure constant 
exceeding the critical value $\alpha_\text{G} \approx 0.5$. Below this value, the Coulomb interaction is too weak to 
create bound states. In this weakly interacting regime, Eqs. (\ref{GBEpol}) and (\ref{GBEpop})
can be solved directly to obtain the optical response and the tight-binding Dirac sea can serve as ground state of the system.
In the regime with strong Coulomb interactions, Eqs. (\ref{GBEpol}) and (\ref{GBEpop}) are not the true graphene Bloch equations, 
since they do not describe the optical response defined with respect to the correct many-body ground state.

\section{Ground State}\label{sec:Groundstate}

\subsection{Derivation of the Gap Equations}

A convenient method to determine the ground state is provided by the variational principle
\begin{equation}
\left.\delta \langle \hat{H}\rangle\right|_{A=0} = 0,
\end{equation}
where $\delta \langle \hat{H} \rangle |_{A=0}$ is the expectation value of the total energy  
in the absence of external fields. 
Due to the Coulomb interaction, the expectation value of the Hamiltonian contains the same many-body correlations 
as the Heisenberg equations of motion and cannot be solved exactly.
To be consistent with the equations of motion, we express the energy expectation value in terms of the same variables
as the equation of motion. Within the mean field approximation, these are $f_{\bk}$, $P_{\bk}$ and $P^*_{\bk}$,
yielding the
variational equation
\begin{equation} \label{deltaH0}
0 = \sum_{\bk} \left( 2\bar\Sigma_\bk \delta \bar f_\bk 
- \bar \Omega^{*}_\bk \delta\bar P_\bk - \bar \Omega_\bk \delta\bar P^*_\bk \right) .
\end{equation}
where barred quantities refer to ground state expectation values.
However, the dynamic variables are coupled and thus cannot be varied independently of each other.

In order to be radiatively stable, the macroscopic current associated with the ground state 
must vanish. This condition is fulfilled for all isotropic distributions $\bar f_\bk$ and $\bar P_\bk$. Due to this 
isotropy, the related Auger contributions to $\bar \Sigma_\bk$ and $\bar\Omega_\bk$ vanish. Furthermore, the ground 
state clearly has to be a stationary state. Therefore, the expectation value of any operator in this state has to be 
static, such that the time derivative of the basic variables has to vanish. Thus, we impose the stationary solution of the 
equations of motion as constraints to assure a steady state,

\begin{eqnarray}
0 &=& 2 \bar\Sigma_\bk \bar P_\bk - (1-2\bar f_\bk ) \bar\Omega_\bk \label{constraintP}\\
0 &=& - 2\Im \left[\bar P^{*}_\bk  \bar \Omega_\bk \right] \label{constraintf}
\end{eqnarray}
Equation (\ref{constraintP}) defines the possible distributions for the nondiagonal populations $\bar P_{\bk}$ in the 
presence of a given isotropic diagonal population $\bar f_{\bk}$ 
subject to the condition that $\bar P_{\bk}$ is real, 
imposed by Eq. (\ref{constraintf}).  In itself, the set of Eqs. (\ref{constraintP},\ref{constraintf}) puts no 
constraints on the populations. As the set of Eqs. (\ref{constraintP}) and (\ref{constraintf}) always has the trivial 
solution $P_\bk=0$, {\it any} isotropic carrier distribution defines at least one
stable equilibrium state with an associated real polarization defined by Eq. (\ref{constraintP}). 
Among these equilibrium states, the ground state minimizes the total energy and can be determined by
inserting Eqs. (\ref{constraintP}) into the variational equation. This yields the condition 
\begin{equation} \label{coherence-condition}
\delta \left[ \bar f_{\bk} (1 - \bar f_{\bk}) - \bar P_{\bk}^2 \right] = 0
\end{equation}
relating $\bar f$ and $\bar P$. 
Assuming that the interacting ground state can be generated adiabatically from the noninteracting 
tight-binding ground state with $f_\bk = P_\bk = 0$,
the combination of the conditional relation and Eq. (\ref{constraintP})
allows us to express the ground state populations in terms of the renormalized energies and Rabi energies:
\begin{eqnarray}
\bar P_\bk &=& \frac{1}{2} \frac{\bar\Omega_\bk} {\sqrt{ \bar\Sigma^2_\bk + \bar\Omega^2_\bk}} \label{eq:gapP}\\
\bar f_\bk &=& \frac{1}{2} \left(1 - \frac{\bar\Sigma_\bk} {\sqrt{\bar\Sigma^2_\bk + \bar\Omega_{\bk}^2 }} \right).\label{eq:gapf}
\end{eqnarray}
Inserting these into Eqs. (\ref{erenorm}) and (\ref{rabi}) yields the coupled set of integral equations:
\begin{eqnarray}
\bar\Omega_{\bk} 
&=& \frac{1}{2} \sum_{\bk'} V({\bk-\bk'}) 
\frac{\bar\Omega_{\bk'}} {\sqrt{\bar\Sigma^2_{\bk'} + \bar\Omega_{\bk'}^2}} \label{eq:gap1} \\
\bar\Sigma_\bk 
&=& \hbar v_F k - \frac{1}{2} \sum_{\bk'} V({\bk-\bk'}) \times \nonumber \\
&\times& \cos(\theta-\theta') \left( 1 - \frac{\bar\Sigma_{\bk'}}{\sqrt{\bar\Sigma^2_{\bk'} 
+ \bar\Omega_{\bk'}^2}} \right) \label{eq:gap2}
\end{eqnarray}
from which the renormalized  and Rabi energies may be calculated numerically. 
The set of Eqs. \ref{eq:gap1} and \ref{eq:gap2} are equivalent to those derived in Ref. \onlinecite{Sabio2010b}
and combines the so called gap-equation for $\bar\Omega_\bk$ with the equation for the renormalized single particle
energy $\bar\Sigma_\bk$.

The set of Eqs. (\ref{eq:gapP})--(\ref{eq:gap2}) displays several interesting features.
From Eq. (\ref{eq:gapf}) one recognizes, that $\bar f_{\bk} > 1/2$ implies a negative 
renormalized energy $\bar\Sigma_{\bk} < 0$ and $\bar f_{\bk} < 1/2$ implies $\bar\Sigma_{\bk} > 0$, respectively. These 
relations can also be deduced directly from equation
\ref{constraintP} by noting that both $\bar P_\bk$ and $\bar\Omega_\bk$ are positive quantities.
From Eqs. (\ref{eq:gap1}) and (\ref{eq:gap2}), we notice the properties
$\bar\Omega_{\bk=0} \geq \bar\Omega_{\bk}$ and $\bar\Sigma_{\bk=0} = 0$
for any isotropic particle distribution, the latter relation resulting from angle integration.  
Hence, for any nontrivial isotropic solution of the gap equation
one has $\bar f_{\bk = 0} = \bar P_{\bk = 0} = 1/2$, i.e. the electrons are in a state 
that mixes the tight-binding valence and conduction bands with equal probability. 

In general, the solution of the gap equation is not unique. This reflects the 
fact that the many-body Hamiltonian may have more than one stationary mean-field solution fulfilling the variational 
condition (\ref{deltaH0}). In particular, the gap equation (\ref{eq:gap1})
always has the trivial solution $\bar\Omega_{\bk}=\bar P_\bk=0$, 
corresponding to a completely incoherent state. 
For the incoherent state, Eq. (\ref{eq:gapf}) simplifies to 
$\bar \Sigma_\bk=\hbar v_F k-\sum_{\bk '}^{k_F} V({\bk-\bk'}) \cos(\theta-\theta')$
with  $\bar f_\bk = (1 - {\rm sg}(\bar\Sigma_\bk))/2 = \theta(k_F - k)$
and $\bar\Sigma_{k_F} = 0$ fixes the Fermi wave number $k_F$. Hence, as $\Sigma_{k=0} = 0$, the tight-binding ground state 
always solves the gap equation and hence corresponds to a stationary mean field solution of the many body Hamiltonian, 
though not necessarily the ground state.

\subsection{Properties of the Ground State}

To achieve intuitive insight into the nature of the ground state, and to derive a criterion for the existence of a nontrivial 
solution of the gap equation, we implement the conditional relation (\ref{coherence-condition}) via
\begin{eqnarray}
\bar f_{\bk} &=& \frac{\beta^2\phi^2(\bk)}{1 + \beta^2\phi^2(\bk)}\label{definephi1}\\
\bar P_{\bk} &=& \frac{\beta\phi(\bk)}{1 + \beta^2\phi^2(\bk)^2}. \label{definephi2}
\end{eqnarray}
Without loss of generality, we can assume $\phi(\bk)$ to be a normalized wave function and $\beta$ 
is an additional parameter controlling the total density.
With the aid of the wave function $\phi$, one can construct the exciton creation and destruction operators
\begin{eqnarray}
\hat B^\dagger = \sum_\bk \phi(\bk) \e^\dagger_{\bk} \h^\dagger_{-\bk}\\
\hat B         = \sum_\bk \phi(\bk) \h_{-\bk} \e_{\bk}
\end{eqnarray}
that generate the transformation $\hat U(\beta) = \exp(\beta \hat B^\dagger)$.  
Acting on the tight-binding ground state, $\hat U$ creates a coherent exciton state
\begin{equation} \label{eq:betastate}
 |\Psi\rangle_{\text{BCS}} = C e^{\beta \hat B^\dagger} |\Psi\rangle _{\text{TB}} \equiv
 \prod_{\bk} \left( u_{\bk} + v_{\bk} \h^\dagger_{-\bk} \e^\dagger_{\bk} \right) |\Psi\rangle_{\text{TB}},
\end{equation}
reproducing the expectation values (\ref{definephi1}) and (\ref{definephi2}). 
Here, the normalization constant is 
$
C = \exp \{ -1/2 \sum_\bk \, \ln(1 + \beta^2\phi^2(\bk) \} .
$ 
The coherent excitonic state is equivalent to the BCS state with
$ u_\bk$ $=$ $1 / \sqrt{1 + \beta^2\phi^2_\lambda(\bk)} $ and
$ v_\bk$ $=$ $\beta\phi(\bk) / \sqrt{1 + \beta^2\phi^2(\bk)}$.

As any Hartree--Fock state is uniquely determined by the expectation value of the basic 
single-particle operators, Eq. (\ref{eq:betastate}) represents the ground state if we identify
$\beta\phi(\bk) = f_\bk/P_\bk$ for any nontrivial solution of the gap equation and fix $\beta$ by normalizing $\phi$.
At low densities, the commutator
\begin{equation}
\left[ \hat B, \hat B^\dagger \right] 
= \sum_{\bk} \phi^2(\bk) \left(1 - \e^\dagger_\bk \e_\bk - \h^\dagger_{-\bk} \h_{-\bk} \right)
\end{equation}
is quasi-Bosonic and the exciton state is formally equivalent to a coherent Glauber state in quantum optics. 

Clearly, for any fixed set $\left\{\phi(\bk)\right\}$, the total particle density is a monotonously 
increasing function of $\beta$ and the tight-binding ground state with 
$\bar P_{\bk} = \bar f_{\bk} = \langle \hat{H} \rangle_{\text{HF}} = 0$ 
corresponds to $\beta = 0$.  
For small $\beta$, the lowest-order contribution to the total energy is given by
\begin{eqnarray}
E(\beta) 
&=& \beta^2 \left[2 \sum_{\bk} \epsilon_\bk \phi(\bk)^2 
 -  \sum_{\bk\bk'} \phi(\bk) V(\bk-\bk') \phi(\bk') \right] \nonumber \\
&+& \mathcal{O}(\beta^4) 
\end{eqnarray}
Subsequent variation of $E/\beta^2 - \mu\sum_\bk \phi(\bk)^2$ yields the linear Wannier equation for the wave 
function $\phi$ with eigenvalue $\mu$. In the limit of vanishing density, the system energy is $E = \beta^2\mu = N\mu$,
where $N = \beta^2$ is the total number of pairs in the low density limit. Hence, if the linear Wannier equation allows 
for bound-state solutions with $\mu < 0$, the system can gain energy by forming bound electron--hole-pair states and a 
nontrivial solution of the gap equation is expected to exist.

\begin{figure}
\centering \includegraphics[width = 8.5cm]{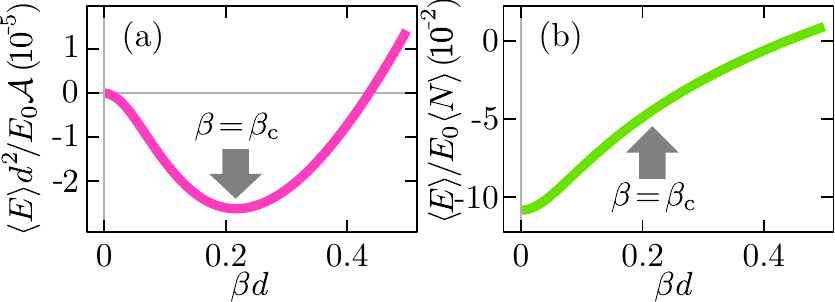}
\caption{Results from evaluating the energy (a) and the energy per particle (b) for the state (\ref{eq:betastate}) 
for different values of $\beta$, with a fixed wave function $\phi(\bk)$.}
\label{fig:beta}
\end{figure}
In Ref. \onlinecite{Gronqvist2011}, bound exciton solutions are reported for parameter conditions corresponding to the regime 
of strong Coulomb interaction. 
These solutions vanish as the system transitions into a weakly interacting regime where the graphene fine-structure constant
$\alpha_\text{G} < 1/2$. To demonstrate the properties of the excitonic transformation (\ref{eq:betastate}), we take the 
wave function $\phi(\bk)$ obtained as numerical solution of the Wannier equation, 
and evaluate the energy of the state (\ref{eq:betastate}) for different values of $\beta$.
For $\alpha_\text{G}=2.4$, which corresponds to the coupling strength in vacuum for vanishing dynamical screening,
the result is shown in the left part of Fig. \ref{fig:beta}.
Increasing  $\beta$ increases the number of pairs contributing to an energy reduction while simultaneously
the energy gain per pair (Fig. \ref{fig:beta}b) decreases due to the Pauli blocking. 
As a result, the obtained energy function has a clear minimum at $\beta_c d\approx 0.2$. We can also note that this 
minimum is clearly below the energy at $\beta = 0$. The energy per pair (Fig. \ref{fig:beta}b) 
at $\beta = \beta_c$ is approximately one half of the exciton binding energy 
$\left.\langle E\rangle/\langle N\rangle\right|_{\beta=0}$.

As any nontrivial solution of the gap equation produces a wave function via $\beta\phi(\bk) = f_\bk/P_\bk$, 
solving the gap equation is equivalent to the minimization of the ground state energy with respect to the 
full set $\beta\phi({\bk})$ and must yield better results than variation with respect to the single 
variational parameter $\beta$. Hence, the existence of bound-state solutions of the linear Wannier equation is 
an unambiguous sign for a nontrivial solution of the gap equation.  

The  solution of the gap equation defines the mean field Hamiltonian 
\begin{eqnarray}
\label{HMF}
\hat{H}^{\text{MF}} \!\!
&=& \sum_{\bk} \bar\Sigma_{\bk} \left( \e^\dagger_{\bk} \e_{\bk} + \h^\dagger_{-\bk} \h_{-\bk} \right)
 - \bar\Omega_{\bk} \left( \h^\dagger_{-\bk} \e^\dagger_{\bk} + \e_{\bk} \h_{-\bk} \right) 
\nonumber\\
\end{eqnarray}
with the single particle dispersion
\begin{equation} \label{eq:newband}
E^{\text{c/v}}_\bk = \pm \sqrt{ \bar\Sigma_\bk^2 + \bar\Omega_\bk^2}.
\end{equation}

The new valence band is below the tight-binding valence band if either $\Omega_\bk=0$ and $\bar\Sigma_{k_F}=0$ has a 
solution with positive Fermi wave number,
in which case the new valence and conduction bands are degenerated at the Fermi level at $k=0$ and $k=k_F$,
or the gap equation has a nontrivial solution with $\bar\Omega_\bk\neq 0$, for which the spectrum exhibits a gap
of magnitude $2\bar\Omega_{\bk = 0}$ at the Dirac points.
Obviously, or rather by construction, the BCS state is the ground state of the mean field Hamiltonian,
and  the energy required to add an electron in the new conduction or valence band is given by $E^{\text{c/v}}_\bk$
and that to create an electron--hole pair by $E^\text{c}_\bk + E^\text{v}_\bk$.
Nevertheless,  the density of
states in the pseudogap is nonzero since
the Bogoliubov state characterized by the order parameters $\left\{\bar\Omega_\bk,\bar\Sigma_\bk\right\}$ 
has a finite overlap to states with order parameters $\left\{\bar\Omega_\bk+\delta\Omega_\bk,
\bar\Sigma_\bk+\delta\Sigma_\bk\right\}$.
Hence, within the Bogoliubov basis, single-particle
excitations are described by scattering within the pseudoparticle
bands, while collective excitations of the many-body system lead to deformation of
the bands that vary the system energy continuously through the
pseudogap.

\subsection{Numerical Solution of the Gap Equations}

\begin{figure}
\centering
\includegraphics[width=8.5cm]{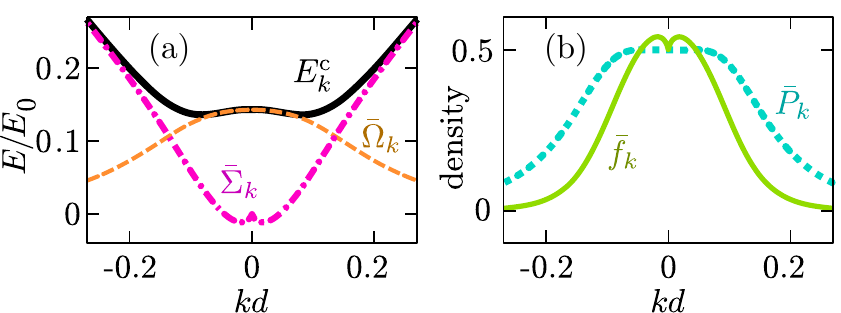}
\caption{
Numerical solutions of the gap equations, Eqs. (\ref{eq:gap1}) and (\ref{eq:gap2}) for $\alpha_\text{G} = 2.4$.
(a) Ground state Rabi energy $\bar\Omega$ (dashed line), ground state renormalized energy $\bar\Sigma$ (dash-dotted line),
and $E^{\text{c}} = (\bar\Omega^2 + \bar\Sigma^2)^{\smash{1/2}}$ (solid line), all versus $kd$.
Energies in units of $E_0 = \hbar v_F / d$.
(b) Ground state population $\bar f$ (solid line) and polarization $\bar P$ (dotted line) versus $kd$. 
These are related to the energies in (a) via Eqs. (\ref{eq:gapP}) and (\ref{eq:gapf}).
Note how both $\bar f$ and $\bar P$ go to 1/2 at $k=0$, and that the region where $\bar f > 1/2$ corresponds to the region 
where $\bar\Sigma < 0$.
} \label{fig:vacuumgapeq}
\end{figure}
Fig. \ref{fig:vacuumgapeq} shows the solution of the full gap equations for 
$\alpha_\text{G}=2.4$. Numerically, we evaluated the gap equation iteratively,
using the optimized excitonic state generated by the solution of the linear Wannier equation 
as initial guess. Convergence was obtained within less than 10 steps. We 
checked the robustness of the numerical solution against the initial guess.
Starting with different values of $\beta$ requires only a few more iterations 
until convergence is reached but produces the same solution.
Due to the form factor in the Coulomb matrix elements, all integrals are well behaved and no additional 
cut-off parameters are required to achieve convergence.
Fig. \ref{fig:vacuumgapeq}a shows the resulting internal field and the renormalized energies
in units of $E_0 = \hbar v_F/d$.
\begin{figure}
\centering
\includegraphics[width=8.5cm]{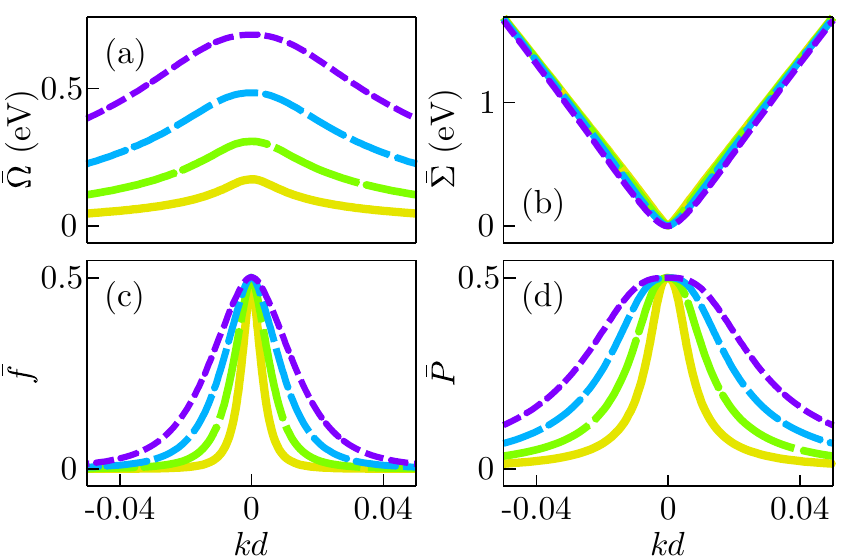}
\caption{
Solutions of the gap equation for different values of the fine-structure constant.
Upper part: (a) ground state Rabi energy $\bar\Omega$ and 
(b) renormalized single particle energy $\bar\Sigma$ versus $kd$ for 
$\alpha_\text{G} = $ 0.9 (solid yellow line), 1.0 (long-dashed green line), 
1.1 (medium-dashed blue line), and 1.2 (short-dashed purple line).
Energy values have been obtained using using $v_F = 9.07\cdot 10^5$ and $d=0.18$\,\AA. 
Lower part: population (c) and polarization (d) distributions.}
\label{fig:alphagapeq}
\end{figure}

We see that the full solution of the gap equation produces a large internal field which decreases monotonously as function of $k$.
Overall, the renormalized energy deviates only slightly from the bare single particle energy, which is due to the angle dependence 
of the Coulomb integrals.
At $k=0$, the renormalized energy starts at zero and for small positive $k$, it becomes slightly negative. In this region 
$\bar f(k) > 1/2$, as can be seen from Fig. \ref{fig:vacuumgapeq}b showing the population and polarization distributions. 
At $k=0$, both $\bar f = 1/2$ and $\bar P = 1/2$. Regarding the first derivative of the renormalized energy as 
renormalized Fermi velocity, our calculations do not confirm  a logarithmic divergence predicted in 
Refs. \onlinecite{Gonzalez99, Juricic2009} but instead, the Fermi velocity decreases and even becomes negative.

In Fig. \ref{fig:alphagapeq}, we show the internal field (Fig. \ref{fig:alphagapeq}a) and renormalized energies 
(Fig. \ref{fig:alphagapeq}b) for $\alpha_\text{G}=$ 0.9, 1.0, 1.1, and 1.2.  
From the Wannier equation, we know that the exciton binding energy decreases rapidly with decreasing effective 
fine-structure constant, and from this, we expect a similar behavior for the internal field.
Indeed, decreasing the coupling constant from $1.2$ to $0.9$ decreases the internal field about one order of magnitude, while the 
renormalized energies are hardly distinguishable from the bare single-particle energies.
Under all conditions, the particle and polarization distribution at $k=0$ are $1/2$ 
see Figs. \ref{fig:alphagapeq} c and d. With increasing wave number, the distributions fall off very rapidly and
the integrated particle and polarization density decrease with decreasing $\alpha_\text{G}$.

\begin{figure}
\centering
\includegraphics[width=8.5cm]{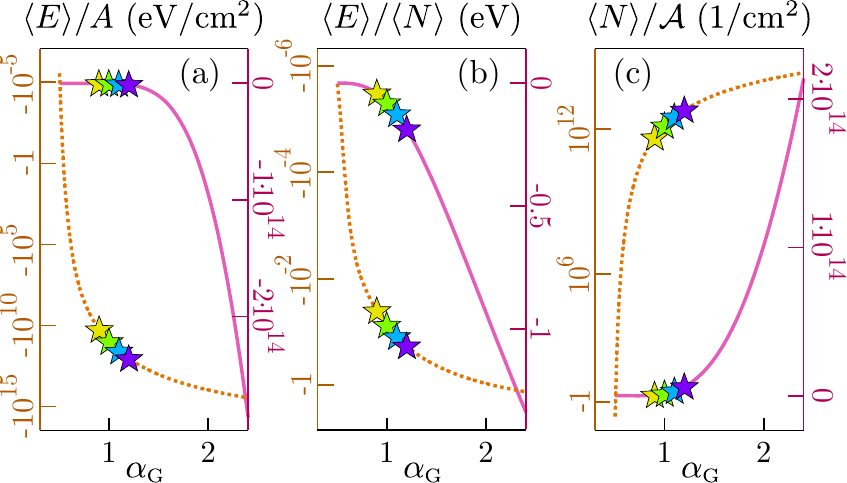}
\caption{Energy density (a), energy gain per pair (b), and pair density (c) as function of the coupling constant $\alpha_\text{G}$.
The dotted curves show the quantities on a logarithmic scale (left axis) and the solid curves present the same quantities 
on a linear scale (right axis). The stars correspond to the gap equations solutions in Fig. \ref{fig:alphagapeq}.
} \label{fig:energydensity}
\end{figure}

In Fig. \ref{fig:energydensity}, we show the total energy density (Fig. \ref{fig:energydensity}a), 
energy gain per pair (Fig. \ref{fig:energydensity}b) and pair density (Fig. \ref{fig:energydensity}c) as function
of
$\alpha_\text{G}$. The energy gain per pair starts at zero for $\alpha_\text{G} = 1/2$ and increases
rapidly with increasing $\alpha_\text{G}$, yielding approximately $1.6$ eV for $\alpha_\text{G} = 2.4$.
Increasing $\alpha_\text{G}$ from $.5$ to $2.4$ increases the total energy gain per ${\rm cm}^2$ over approximately 
20 orders of magnitude, which is a combined 
effect of the increasing energy gain per pair and the increasing pair density. At $\alpha_\text{G}=2.41$ the pair density is
$\langle N \rangle /\mathcal{A} = 2.16 \cdot 10^{14}{\rm cm}^{-2}$, which should be compared
with the total density $n_0 = 1.27 \cdot 10^{15} {\rm cm}^{-2}$ of the valence band electrons.

\begin{figure}[ht]
\centering
\includegraphics[width=8.5cm]{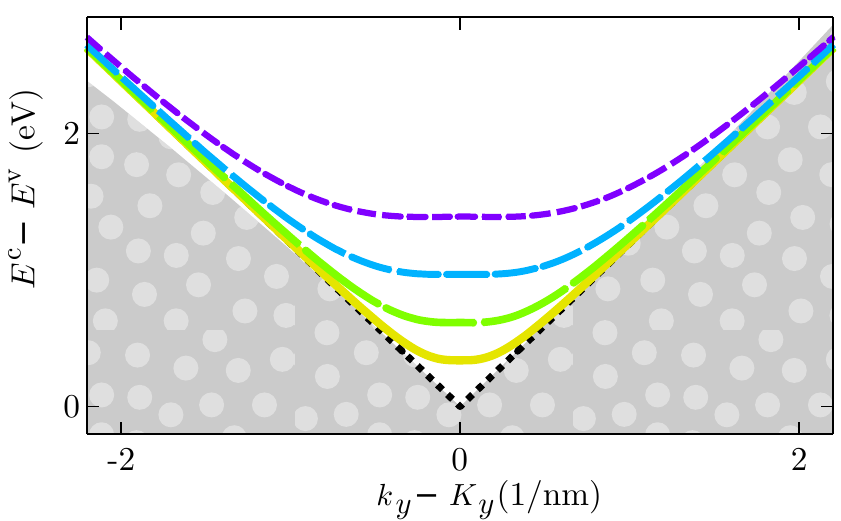}
\caption{
Dependence of the quasiparticle band structure $E^{\text{c}}_\bk-E^{\text{v}}_\bk$ (\ref{eq:newband})
on the effective fine-structure constant $\alpha_\text{G}$.
The curves shown are for $\alpha_\text{G} = $ 0.9 (solid yellow line), 1.0 (long-dashed, green line), 
1.1 (medium-dashed, blue line), and 1.2 (short-dashed, purple line), 
together with the linear tight-binding dispersion (dotted black line). 
The shaded, dotted area shows the full tight-binding dispersion.
These band structures correspond to the gap equations solutions in Fig. \ref{fig:alphagapeq}.}
\label{fig:alphabands}
\end{figure}
Fig. \ref{fig:alphabands} shows the resulting quasiparticle dispersion for different values of the effective fine-structure 
constant in comparison with the noninteracting TB bands. For values of $\alpha_\text{G}$ exceeding the critical value
$\alpha_c=1/2$, a gap opens that increases with with increasing $\alpha_\text{G}$. For values of $\alpha_\text{G}<1.1$,
the Bogoliubov bands lie well within the validity range of the linear approximation to the tight-binding bands. 
For higher values, corrections beyond the linear approximation should be taken into account, 
however, this is beyond the scope of this article.

\section{Optical Response} \label{sec:GBE}
To illustrate the influence of Coulomb effects on measurable quantities, we calculate the optical response of our graphene 
model system. For this purpose, we derive the graphene Bloch equations (GBE) and compute the linear response to an external 
optical field. In all calculations, the initial conditions will be chosen such that the graphene sheet is excited from the BCS 
ground state by a normally incident optical or terahertz pulse. 

As shown in section \ref{sec:Hamiltonian}, at normal incidence, the two degenerate Dirac points couple to the right and left 
circular polarized components of the transverse optical field. 
Hence, the circular polarization components provide a natural basis for the optical field in which the different polarizations
are decoupled. Each polarization component of the vector potential obeys the wave equation
\begin{equation}
\label{maxwell}
\Delta { A^\pm}-\frac{\epsilon}{c^2}\frac{\partial^2}{\partial t^2}{ A^\pm}=-\frac{4\pi}{c}{ j^\pm}.
\end{equation}
Here, $j^\pm=-c\langle\frac{\delta H_I}{\delta A^{\pm*}}\rangle$ is the expectation value of the total current
\begin{eqnarray}
\ j^\pm
& = & -\frac{e^2n_0}{m_0c} A^\pm(z=0,t) f(z) \nonumber\\
& + & \frac{e|{\bm \pi}|}{m_0} \sum_{\bk} \biggl\{ (1-2f_\bk) \cos\theta
\mp \Im \left[ P_\bk\right] \sin\theta \biggr\} f(z) \nonumber\\
& = & - \frac{e^2n_0}{m_0c} A^\pm(z=0,t) f(z) + j^\pm_{\left[p\right]}f(z) \label{current}
\end{eqnarray}
with $f(z)=\int d^2\rho|\phi_{p_z}({\bm r})|^2\approx\delta(z)$ on the scale of the optical or terahertz wavelength. 
The total current consists of two contributions. The first is simply proportional to the vector potential 
and hence purely classical.
Its only dependence on material quantities is via the charge density $\rho({\bm r})=-e n_0f(z)$, where $n_0$ is the 
total density of the electrons in the $\pi$-band.
This term is responsible for a Drude-like response. 
The second part of the current is purely particle like and associated with 
inter-  and intraband transitions. Due to the phase $\theta$, isotropic electron and hole distributions or 
polarizations do not contribute to a macroscopic particle current. 
Hence, if the system is in its ground state and in the absence of external fields, there is no macroscopic current.

\subsection{Graphene Bloch Equations}

To calculate the particle current, we divide the polarization and the particle distributions  
into a static part arising from the ground state, and a dynamical part arising from the optical excitations:
\begin{eqnarray}
 P_{\bk} &=& \bar P_{\bk} + \Delta P_{\bk},\\ 
 f_{\bk} &=& \bar f_{\bk} + \Delta f_{\bk}
\end{eqnarray}
and correspondingly for the renormalized and Rabi energies. 
In a first approximation, all contributions beyond Hartree--Fock are treated phenomenologically by introducing constant 
dephasing and relaxation rates, i.e. $\frac{\ud}{\ud t} \Delta P_{\bk}=-\gamma_2\Delta P_{\bk}$ and 
similar for the populations.
Noting that $ \bar P_{\bk}$ is real, the GBE for the dynamical part of the polarization 
and particle distribution are obtained as
\begin{eqnarray}
i\hbar \frac{\ud}{\ud t} \Delta P_{ \bk}
&=& 2\left(\bar\Sigma_{\bk} + \Delta \Sigma^r_{\bk}\right)\Delta P_{\bk}
   - (1 - 2\bar f_{\bk } - 2\Delta f_{\bk }) \Delta\Omega^R_{\bk} \nonumber\\
&+&  2\Delta\Sigma_{\bk}\left( \bar P_{ \bk}+\Delta  P_{\bk}\right)
   + i (1 - 2\bar f_{\bk } - 2\Delta f_{\bk }) \Delta\Omega_{\bk}\nonumber\\
&+&  2\bar P_{ \bk}\Delta\Sigma^r_{\bk}
   + 2\bar\Omega_{\bk}\Delta f_{\bk }
   - i\hbar\gamma_2\Delta  P_{ \bk}, \label{GBEpol2}\\
\hbar \frac{\ud}{\ud t}\Delta f_{\bk} 
&=& - 2\Im\left[\Delta P_{ \bk}^*\Delta\Omega^R_{\bk}\right] \nonumber\\
&+& 2\Delta\Omega_{\bk}\left(\bar P_{\bk} + \Re\left[\Delta P_{ \bk}\right]\right) \nonumber\\
&-& 2\Im\left[\bar P_{ \bk}\Delta\Omega^R_{\bk}+
    \bar\Omega_{\bk} \Delta P^*_{ \bk} \right]
  -  \hbar\gamma_1 f_\bk .\label{GBEpop2}
\end{eqnarray}

In Eq. (\ref{GBEpol2}), the first line is the direct generalization of the familiar homogeneous part of the SBE,
producing the generalized Wannier equation. The dynamical part of the Rabi energy $\Delta\Omega^R_\bk$ contains the optical
field and the internal field of the optically induced interband polarization only, while both the ground state and the dynamical
populations contribute to the renormalization of the single particle energies and the phase space filling, 
reducing the effective Coulomb interaction.
The second line describes an energy renormalization due to the optical field and Auger scattering  proportional 
to $\Delta\Sigma_\bk$, and Auger contributions proportional to $\Delta\Omega_\bk$ that act as a source/drain for the 
dynamical polarization. These processes are very ineffective in a wide gap semiconductor and are usually 
neglected in the SBE. The last line Eq. (\ref{GBEpol2}) describes the coupling of the polarization to the populations via the 
ground state polarizations. These contributions do not exist in the SBE and are specific for an excitonic ground state.

The first line in the equation of motion for the populations, Eq. (\ref{GBEpop2}), is the direct generalization of the SBE. 
The respective terms contain dynamical quantities only and are at least of second order in the optical field. The second line 
describes the conversion of polarizations into populations via Auger scattering processes proportional to
$\Delta\Omega_\bk $. The last line describes ground state polarization assisted sources, specific for the excitonic ground state.

The major effect of the ground state polarization is the introduction of a linear source for the optically 
induced populations $f_\bk$. As a result, within a power expansion in terms of the exciting field, both the dynamical 
polarization and the population contain all powers of the exciting field. For any arbitrary order, the equations of motion 
are inherently coupled by the contributions proportional to $\bar P_{\bk}$ in the last lines of Eqs. (\ref{GBEpol2}) and 
(\ref{GBEpop2}), and hence, must be solved simultaneously rather than iteratively.
It is exactly this polarization mediated coupling that is responsible for the occurrence of the Bogoliubov gap in the 
linear spectrum.

\subsection{Linear Optical Spectra}

\begin{figure}
\vspace{.5cm}
\centering{\includegraphics{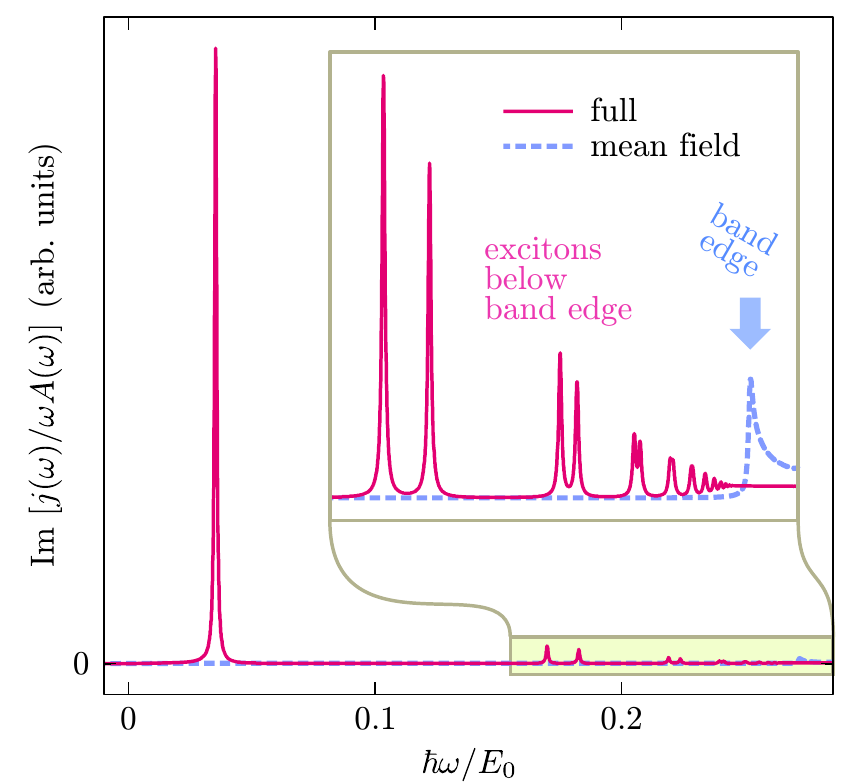}}
\caption{Imaginary part of the linear susceptibility for a graphene-like system with $\alpha_\text{G}=2.4$. 
Energy in units of $E_0 = \hbar v_F / d$.}
\label{fig:vacuumspectrum}
\end{figure}

To illustrate the basic effects of several contributions to the optical response,
we solve the linearized GBE for a given external optical field. 
Examples of the results for $\alpha_\text{G}=2.4$  and $\hbar\gamma = 5\cdot10^{-4} E_0$ are shown in Fig. \ref{fig:vacuumspectrum}. 
The spectrum shows $4\pi\Im[j^\pm/\omega A^\pm]$ which is a direct measure for the absorption.
We notice pronounced excitonic resonances at low energies followed by a spectrally flat response.
Due to the optical selection rules, all bright excitonic resonances have a $p$-like symmetry.
Similar to the spectra in semiconductors, the peak height increases with increasing binding energy, 
showing that oscillator strength is transferred to strongly bound excitons.

Ignoring all Coulomb effects in the GBE, other than those responsible for the excitonic ground state, gives the
artificial spectrum shown by the dashed (blue) curve in Fig. \ref{fig:vacuumspectrum}. 
This solution is equivalent to that obtained using only the mean-field Hamiltonian (\ref{HMF}). The resulting spectrum 
is proportional to the density of states (DOS) (divided by $\omega$) of the Bogoliubov bands and has no additional structure.
The onset of the continuum absorption starts with a $1/\sqrt{\hbar \omega-E_G}$-like 
damped singularity exactly at the Bogoliubov gap. The origin of the singularity is the shift of the band minimum to finite 
$k$-values (see Figs. \ref{fig:vacuumgapeq} and \ref{fig:alphagapeq}). Within a quadratic approximation 
$E(k) = E_G+\hbar^2(|k|-k_0)^2/2m_{\rm eff}$, the DOS 
\[
g(E) = \frac{m_{\rm eff}}{\pi\hbar^2}\theta(E-E_G)\left(1+\sqrt{\frac{\hbar^2k_0^2}{2 m_{\rm eff}(E-E_G)}}\right), 
\]
consists of a step-like contribution typical for 2D parabolic bands and a singular part proportional to the shift $k_0$ that is
similar to the free-particle result of a semiconductor quantum wire.\cite{HaugKoch}
Energetically above the Bogoliubov gap, the optical response assumes the 'universal' value given by 
$\frac{1}{2}\pi\alpha(e^*/e)^2$ where $\alpha = e^2/\hbar c$ 
is the fine-structure constant and  $e^* = e|{\bm \pi}|/\sqrt{2}m_0v_F$ is the effective charge, producing the effective medium 
Hamiltonian $H_0 + \smash{H_I^{\left[p\right]}} = v_F{\bm \sigma}\cdot \left(\bp - e^*/c{\bm A} \right)$ 
where ${\bm \sigma}=(\sigma_x, \pm\sigma_y)$ combines the Pauli spin matrices. 
The effective charge depends on the optical matrix element and can either be calculated from the carbonic wave 
functions or be used as a fit parameter. 
An additional factor 2 arises if the spin degeneracy is taken into account. 
Note that there is no valley degeneracy for the excitation with circularly polarized light.

Including all Coulomb interaction terms, we obtain several clearly recognizable excitonic resonances below the pseudogap, 
shown by the solid (pink) curve in Fig. \ref{fig:vacuumspectrum}. The resonances arise from the Coulomb interaction of 
the optically induced excitations only, while the excitonic ground state populations open the required gap. 

\begin{figure}
\vspace{.5cm}
\centering{\includegraphics[width=8cm]{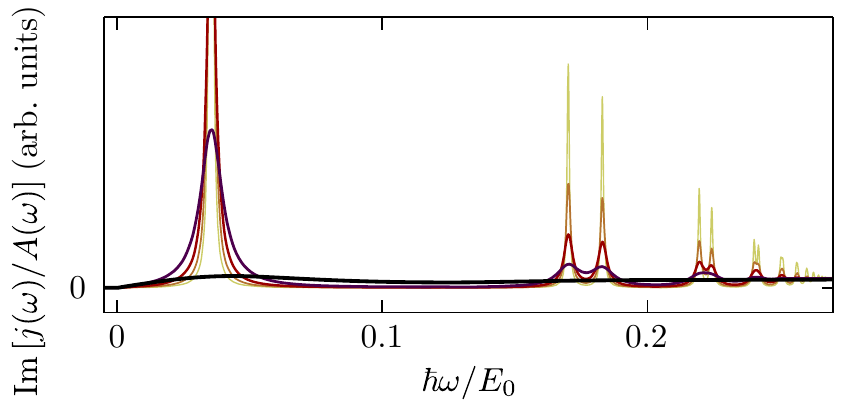}}
\caption{Effect of dephasing on the spectra of a graphene-like system with $\alpha_\text{G}=2.4$. 
The curves show the results for dephasing constants (from most to least peaked) 
$\hbar \gamma/E_0 = 5 \cdot 10^{-4}$, $1 \cdot 10^{-3}$, $2\cdot 10^{-3}$, $5 \cdot 10^{-3}$, and $5 \cdot 10^{-2}$.
Energies in units of $E_0 = \hbar v_F / d$.}
\label{fig:spectrumdephasing}
\end{figure}
To study the influence of dephasing, we repeated the calculation for the dephasing values in the range from  $\hbar \gamma
= 5 \cdot 10^{-4} E_0$ (lightest, most peaked curve) to  $\hbar \gamma = 5 \cdot10^{-2} E_0$ (darkest, flattest curve). 
As we can see in Fig. \ref{fig:spectrumdephasing} the excitonic spectra broaden and the fine structure close to the gap 
smoothens to a flat, continuum-like response. The lowest exciton resonance is clearly recognizable for dephasing rates
$\hbar\gamma$ up to $5 \cdot 10^{-3} E_0$.

\begin{figure}
\vspace{.5cm}
\centering{\includegraphics{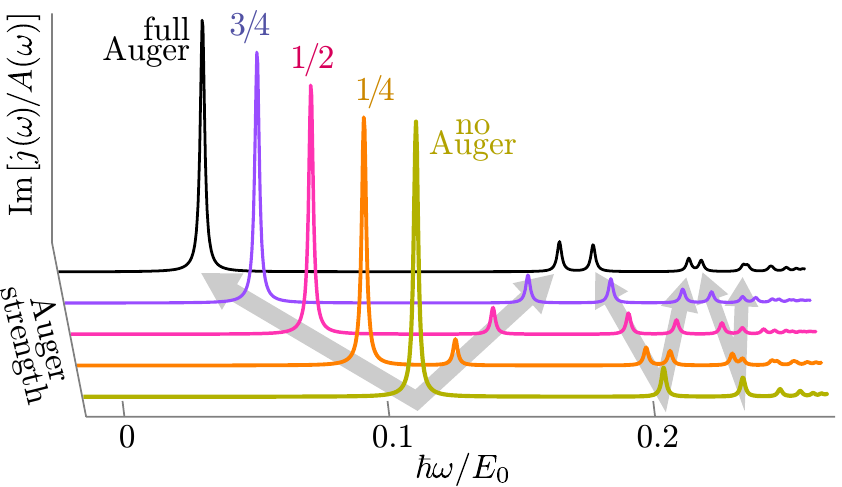}}
\caption{Imaginary part of the linear susceptibility for a graphene-like system with $\alpha_\text{G}=2.4$. 
Energy in units of $E_0 = \hbar v_F / d$.}
\label{fig:spectrumAuger}
\end{figure}

To study the influence of the Auger contributions, we present in Fig. \ref{fig:spectrumAuger} results where we 
artificially vary the relative strength of the Auger terms between 0 (no Auger terms) and 1 (full Auger contributions).
The comparison shows that the Auger contributions not only increase the exciton binding energy significantly,
but they are also responsible for the splitting of the individual resonances. 
In the absence of the Auger contributions, the two distinct projection states of the angular momentum onto the 
propagation direction of the incident light are degenerated. As the Auger contributions invert the rotational symmetry 
they remove this degeneracy. The splitting of the resonances increases with increasing strength of the Auger terms
and is symmetric with respect to the energy shift. The oscillator strength
is distributed nonequally among the resonances. The curves in Fig. \ref{fig:spectrumAuger} show $\Im\left[j/A\right]$, which 
differs by a factor $1/\omega$ from the absorption spectra. This presentation  shows that the the heights of the absorption 
peaks vary exactly like $1/\omega$, with a constant prefactor that is considerably larger for the lowest lying state and independent 
of the strength of the Auger terms. As the Auger terms vanish for $s$-like states, the splitting confirms the $p$-like 
symmetry of the optically bright states.

\begin{figure}[t]
\vspace{.5cm}
\centering{\includegraphics{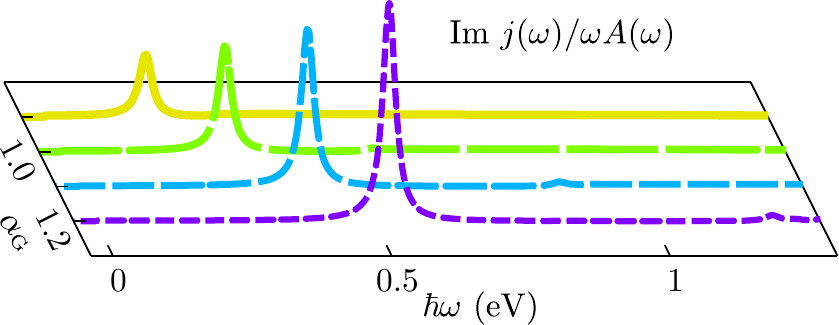}}
\caption{Imaginary part of the linear susceptibility for a graphene-like system for 
$\alpha_\text{G} = 0.9$ (solid yellow line), $\alpha_\text{G} = 1.0$ (long-dashed green line),
$\alpha_\text{G} = 1.1$ (medium-dashed blue line) and $\alpha_\text{G} = 1.2$ (short-dashed purple line).
These spectra correspond to the gap equations solutions in Fig. \ref{fig:alphagapeq}
and to the band structures in Fig. \ref{fig:alphabands}.
}
\label{fig:spectrumalphas}
\end{figure}

In Fig. \ref{fig:spectrumalphas} the absorption spectra for several values of the coupling strength and a fixed dephasing 
rate $\hbar \gamma = 13$ meV are shown. 
Variation of the effective coupling strength can be realized by embedding the graphene sheet in, or putting it on top of a 
dielectric medium, altering the static screening $\epsilon=\epsilon_r$ or $\epsilon=(1+\epsilon_r)/2$ respectively.
In the strong Coulombic regime with $\alpha_\text{G}>1/2$, all spectra show one clearly recognizable discrete absorption peak and 
a flat continuum response. Both the spectral positions and the height of the absorption peak depend extremely sensitive
on the effective coupling strength, i.e. on screening by the dielectric environment.
Qualitatively, any single spectrum is similar to that of a conventional semiconductor. However, the dependence on 
the coupling strength is very different. 
If one compares e.g. quantum well systems with a different background screening and defines the effective 
coupling strength as
$\alpha_{\rm eff}=e^2/\epsilon\hbar c=\alpha/\epsilon$, the spectral position of the gap is 
independent of the background refractive index while the exciton binding energy varies like 
$\alpha_{\rm eff}^2 m_{\rm r}c^2$ and $m_{\rm r}$ is the reduced mass of the electron--hole pair.
The net result is a red shift of the excitonic resonances with increasing coupling strength.
In graphene, the exciton binding energy, defined with respect to the Bogoliubov band edge, increases also with 
increasing coupling strength, but simultaneously, the band edge shifts towards higher energies, resulting in a net
{\it blue} shift of the lowest excitonic resonance with increasing coupling strength.

\section{Summary and Conclusions}

In conclusion, we presented a framework to determine the ground state and optical response of graphene and graphene-like 
systems.
Our method is based on the equations of motion for the basic variables, combined with a variational Ansatz for the ground 
state. Even though we have only presented results in linear approximation,
our method allows for a selfconsistent inclusion of e.g. higher order many-body correlations or dynamical screening,
treating the ground state properties and excitation dynamics on equal footing. 
It can easily be generalized to finite temperature, be adapted to other strongly correlated systems 
or to include additional constraints like doping.

Within the Hartree--Fock approximation, our procedure produces the gap equation for the ground state.
We could relate the criterion for exciton condensation of the ground state with the existence of bound 
exciton solutions of the linear Wannier equation discussed in Ref. \onlinecite{Gronqvist2011} and
solved the gap equations numerically. Our results for the ground state are in general agreement with
 both Monte Carlo simulations and variational approaches studying the semimetal to insulator 
transition\cite{Sabio2010b, Gamayun2010} and predict a 
a critical value of $\alpha=0.5$ for the onset of exciton condensation.

Consistently with the level of approximation for the ground state, we derived the graphene Bloch equations for the 
strongly interacting regime on the singlet level, which allows us to calculate the optical response from the BCS ground state.
Unlike in direct-gap semiconductors, where electron--hole recombinations are dipole allowed, $s$-like excitons in 
graphene are optically dark and hence radiatively stable. 
Signatures of the excitonic ground state are the existence of a pseudo gap and several bright, 
$p$-like excitonic resonances  in the optical spectra, with a spectral position depending extremely 
sensitively on the effects of screening.

Topics of future research are e.g. the inclusion of dynamical screening, 
that might reduce the nominal effective fine-structure constant $\alpha_\text{G}=2.4$\cite{Reed2010} significantly,
the inclusion of Coulomb scattering, correlations beyond the singlet level, as well as optical nonlinearities.

\raggedright

\end{document}